\documentclass[12pt]{article}
 \pdfoutput=1
\usepackage{amsfonts,amsthm,amsmath,amssymb,slashed}
\usepackage[textwidth = 500 pt, textheight = 630 pt]{geometry}
\usepackage{latexsym,amsfonts,amsmath,amssymb,theorem,dsfont,wasysym}
\usepackage{amssymb}
\usepackage{amsmath}
\usepackage{amstext}
\usepackage{graphicx,epsfig}
\usepackage{epsfig}
\usepackage{verbatim} 
\usepackage{fancyhdr}
\usepackage{fancybox}
\usepackage{color}
\usepackage{ulem,bbold}
\usepackage{enumitem}
\usepackage{subfigure}
\usepackage{bbm}
\usepackage{parskip}
\usepackage[numbers,sort&compress]{natbib}

\definecolor{MyDarkBlue}{rgb}{0.15,0.15,0.45}
\usepackage[linktocpage=true]{hyperref}
\hypersetup{
colorlinks=true,
citecolor=MyDarkBlue,
linkcolor=MyDarkBlue,
urlcolor=MyDarkBlue,
}

\usepackage[numbers,sort&compress]{natbib}

\def\beq{\begin{eqnarray}}
\def\eeq{\end{eqnarray}}

\def\({\left(}
\def\){\right)}

\newcommand{\be}{\begin{equation}}
\newcommand{\ee}{\end{equation}}

\def\ea{\end{eqnarray}}
\def\ba{\begin{eqnarray}}

\def\beq{\begin{eqnarray}}
\def\eeq{\end{eqnarray}}

\def\({\left(}
\def\){\right)}

\def\lsim{\mathrel{\rlap{\lower3pt\hbox{\hskip0pt$\sim$}}
     \raise1pt\hbox{$<$}}}         
\def\gsim{\mathrel{\rlap{\lower4pt\hbox{\hskip1pt$\sim$}}
     \raise1pt\hbox{$>$}}}         

\def\lsim{\mathrel{\rlap{\lower3pt\hbox{\hskip0pt$\sim$}}
     \raise1pt\hbox{$<$}}}         
\def\gsim{\mathrel{\rlap{\lower4pt\hbox{\hskip1pt$\sim$}}
     \raise1pt\hbox{$>$}}}         

\usepackage{amsmath}
\usepackage{amsfonts}
\usepackage{verbatim}
\usepackage{graphicx}
\usepackage{subfigure}
\usepackage{amssymb}
\usepackage[T1]{fontenc}
\usepackage{slashed}

\begin{document}

\renewcommand{\thefootnote}{\fnsymbol{footnote}}

\makeatletter
\@addtoreset{equation}{section}
\makeatother
\renewcommand{\theequation}{\thesection.\arabic{equation}}

\rightline{}
\rightline{}



\begin{center}
{\Large \bf{On Effective Theory of Superfluid Phonons}}

 \vspace{1truecm}
\thispagestyle{empty} \centerline{\large  {Lasha  Berezhiani}
}

 \textit{Max-Planck-Institut f\"ur Physik, \\ F\"ohringer Ring 6, 80805 M\"unchen, Germany\\ \vskip 10pt
 Arnold Sommerfeld Center, Ludwig-Maximilians-Universit\"at, \\Theresienstra{\ss}e 37, 80333 M\"unchen, Germany
  }

\end{center}  
 
\begin{abstract}

It is shown that the standard effective field theory of non-relativistic superfluids, which includes only the leading order action in derivative expansion, is in general inadequate for studying supersonic processes. Namely, for a proper treatment of the latter the phonons with momenta greater than $mc_s$ are relevant at times (with $m$ being the mass of the superfluid constituent and $c_s$ denoting the sound-speed), which lie beyond the regime of validity of the naive effective theory mentioned above. This is demonstrated on a concrete example by recovering the known expression for the  friction force as a process of radiating phonons by a supersonic perturber. In particular, when the impurity is heavier than superfluid constituents, it is indeed necessary to extend the non-relativistic effective theory of superfluids to include high-derivative corrections. For the light perturber, however, there is a window of velocities for which high-derivative corrections are irrelevant.

\end{abstract}

\newpage
\setcounter{page}{1}

\renewcommand{\thefootnote}{\arabic{footnote}}
\setcounter{footnote}{0}

\linespread{1.1}
\parskip 4pt

\section{Introduction}

Much attention has been devoted to understanding the effective field theory formalism for superfluid phonons and its employment for studying various phenomena relevant for cosmology and astrophysics \cite{Greiter:1989qb,Son:2002zn,Son:2005rv,Nicolis:2011cs,Nicolis:2017eqo}. Besides it being an interesting substance with interesting properties on its own, that can be manufactured in laboratories from known particles, superfluids have been theorised to be potentially relevant even in dark matter context. In particular, in some scenarios dark matter may undergo superfluid phase-transition cosmologically and/or in localised structure \cite{Goodman:2000tg,Slepian:2011ev,Guth:2014hsa,Berezhiani:2015pia,Berezhiani:2015bqa,Alexander:2016glq,Berezhiani:2017tth,Sharma:2018ydn,Alexander:2019qsh,Berezhiani:2019pzd}. In some other cases, the superfluid He has been proposed as a target material for dark matter detection \cite{Knapen:2016cue,Hertel:2019thc,Acanfora:2019con,Caputo:2019cyg}. The effective field theory of superfluids is a central tool for some of these considerations. 

The goal of this letter is to explore the limitations of the widely discussed effective field theory, which is usually terminated at the leading order in derivative expansion (so-called $P(X)$ theory). We demonstrate the frequently overlooked point (within cosmology literature) regarding supersonic processes; namely, that such processes may lie beyond the sound-wave regime and the extension of the effective theory to include high-derivative corrections is warranted. The same point has been recently shown to apply to the case of gravity-mediated drag force \cite{Berezhiani:2019pzd}. The proper calculation of the superfluid friction force has been previously carried out in \cite{Pitaevskii} by solving the Gross-Pitaevskii equation in the presence of the impurity.

\section{Theory and background}
\label{superfluid}

We consider a self-interacting massive complex scalar field $\Phi$, which can form a superfluid upon condensation. In addition, we consider a real scalar field $\chi$, which will play the role of an impurity.  The Lagrangian density we focus on, which has these properties, is given by
\beq
\mathcal{L}=-|\partial\Phi|^2-m^2|\Phi|^2-\frac{\lambda}{2}|\Phi |^4-\frac{1}{2}(\partial\chi)^2-\frac{1}{2}M^2\chi^2-\frac{1}{2}g\chi^2|\Phi|^2\,.
\label{lag}
\eeq
For the stability of the homogeneous condensate we need $\lambda>0$ corresponding to the repulsive self-interactions, which we will be assuming throughout the paper. This is the simplest theory for the top-down analysis of superfluidity, and that is precisely the reason we are focusing on it.

The classical background corresponding to the condensate of interacting $\Phi$'s is given by
\beq
\Phi=ve^{i \mu t}\,,\qquad \chi=0\,, \qquad \text{with} \qquad \mu^2=m^2+\lambda v^2\,.
\label{superfluidsoln}
\eeq
It is straightforward to show that this solves the equations of motion following from \eqref{lag}. The particle number density in the condensate, defined as the $U(1)$ charge density, corresponding to \eqref{superfluidsoln} reduces to $n=2\mu v^2$.

In quantum mechanical description, the classical configuration given by \eqref{superfluidsoln} corresponds to the degenerate state of large number of $\Phi$ particles in zero momentum state, while the occupation number of $\chi$ particles vanishes. Notice that if condensate constituents were non-interacting ($\lambda=0$), then the background would have been $\Phi=ve^{imt}$ and correspond to ideal BEC. As it is well known, and as we are about to see, the presence of self-interactions is vital for the condensate to exhibit superfluidity.

The condensate state \eqref{superfluidsoln} spontaneously breaks a diagonal combination of the time-translation and $U(1)$ symmetry, consequently we expect a Nambu-Goldstone boson in the spectrum. In fact, it is straightforward to show that out of two degrees of freedom, residing in $\Phi$, one is heavy with mass $2 m$, assuming $\lambda v^2\ll m^2$, and the other is massless. At low energies the massive mode becomes nondynamical and the effective theory of the massless degree of freedom can be obtained by integrating out the heavy mode. Moreover, the mass of $\chi$ gets shifted in the presence of the condensate and becomes $m_\chi^2=M^2+g v^2$. For simplicity we will work in the approximation $m_\chi\approx M\gg \sqrt{g}v$.

\section{Mean-field description of superfluidity}
\label{superfluidmean}

To study perturbations around the fixed classical background \eqref{superfluidsoln}, let us parametrise fluctuations in the usual fashion
\beq
\Phi=(v+\rho)e^{i(\mu t+\pi)}\,.
\label{supdec}
\eeq
As pointed out, we expect only one massless excitation, commonly referred to as 'phonon'. In order to derive the effective field theory for this one massless degree of freedom and its interactions with $\chi$ field, we substitute \eqref{supdec} into \eqref{lag} and take the non-relativistic limit; resulting in
\beq
\mathcal{L}=-(\partial_j \rho)^2+2 m (v+\rho)^2\left\{ \frac{\lambda v^2}{2m}+\dot{\pi}-\frac{(\partial_j \pi)^2}{2 m}-\frac{g}{4m}\chi^2 \right\}-\frac{\lambda}{2}(v+\rho)^4\,;
\label{nrsuplag}
\eeq
with the kinetic and mass terms for $\chi$ being implicit.
Here, the non-relativistic limit corresponding to $\partial \ll m$, together with $\lambda v^2\ll m^2$ which is equivalent to having a subsonic sound-speed for phonons.
As can be easily seen from \eqref{nrsuplag}, the field $\rho$ is non-dynamical, i.e. its equations of motion do not involve its time derivatives. Therefore it can be integrated out, see the appendix for the derivation of the effective theory of phonons. In order to find the dispersion relation of a superfluid phonon -- the single dynamical degree of freedom in $\Phi$-sector at low energies -- one can linearise \eqref{nrsuplag} and analyse the characteristic determinant to find
\beq
\omega_k^2=c_s^2 k^2+\frac{k^4}{4m^2}\,, \qquad \text{where} \qquad c_s^2\equiv \frac{\lambda n}{4m^3}\,.
\label{disp}
\eeq
The first term in the above-given dispersion relation corresponds to the sound wave; while the second term corresponds to the propagation of the massive particle. In other words, small-momentum phonons propagate like waves with speed $c_s$, while the energy cost for producing phonons with high-momenta -- for which the second term of \eqref{disp} dominates -- is the kinetic energy for a massive particle. 

The effective theory for a phonon can be obtained by integrating out $\rho$-field from \eqref{nrsuplag}. As a result we will get high-spatial-derivative theory for $\pi$, giving rise to the dispersion relation given by \eqref{disp}. However, for the clarity of our little demonstration let us begin the discussion with the phonon theory at low momenta. In particular, let us derive the effective theory of phonons for $k\ll m c_s$, i.e. for momenta for which the first term in \eqref{disp} dominates. In other words, we would like to derive effective theory of phonons in the leading order in derivative expansion. For this, we ignore the gradient term for $\rho$-field which results in the Lagrangian where $\rho$ enters algebraically. After integrating it out we arrive at the following Lagrangian describing the phonon dynamics
\beq
\mathcal{L}=\frac{(2m)^2}{\lambda}\left( m c_s^2 +\dot{\pi}-\frac{(\partial_j \pi)^2}{2 m}-\frac{g}{4m}\chi^2\right)^2\,.
\label{paction}
\eeq
Here, we do not write the kinetic term for $\chi$ explicitly, but it is implied. This Lagrangian describes phonon dynamics to the leading order in derivative expansion for a special superfluid whose pressure originates from the repulsive 4-point contact interaction; and is a representative of the class of effective theories for generic superfluids \cite{Greiter:1989qb,Son:2002zn,Son:2005rv} (supplemented with an additional spectator field $\chi$). Notice the emergence of the effective 4-point interaction for $\chi$ with the coupling strength given by $g^2/\lambda$. Obviously, the validity of the perturbative description demands this effective coupling to be small. In other words, the point to emphasize here is that \eqref{paction} breaks down in $\lambda\rightarrow 0$ limit and the inclusion of high-derivative interactions is mandatory, see \cite{Berezhiani:2018oxf}.

The relevant part of the Lagrangian for computing the energy dissipation rate for moving $\chi$-particle can be obtained from \eqref{paction} straightforwardly; after canonical normalisation  $\pi\rightarrow \pi \sqrt{\lambda}/2m$, reducing to
\beq
\mathcal{L}=\frac{1}{2}\left( \dot{\pi}^2-c_s^2 (\partial_j \pi)^2 \right) -\frac{g}{2\sqrt{\lambda}}\chi^2\dot{\pi}+\ldots \,,
\label{plag}
\eeq
where ellipses indicate interaction terms nonlinear in the phonon field. Neglecting multi-phonon processes corresponds to working within linear response approximation, which is the limit the calculation of \cite{Pitaevskii} was done for.

As we see, in the approximation at hand, the dispersion relation of $\pi$ is given by
\beq
\omega_k=c_s k\,.
\label{sounddisp}
\eeq
Using this, it is easy to show that the dissipation process is kinematically prohibited unless the motion of the $\chi$-particle is supersonic. Assuming that this condition is satisfied, it is straightforward to compute the energy dissipation rate. However, we are about to show that the approximation \eqref{plag} is in general inadequate for obtaining the correct dissipation rate. \footnote{See \cite{Nicolis:2017eqo} for the calculation of some of the scattering rates within the effective theory of a generic superfluid to the leading order in derivative expansion; i.e. within the framework similar to \eqref{paction}.}

The rate of the process $\chi\rightarrow \chi+\pi$ can be computed using\footnote{We are interested in finding the parametric dependence of the rate on physical quantities, as such we will be dropping the overall numerical factors throughout this work.}
\beq
d\Gamma= \frac{1}{E_{\chi}^{in}} \left( \frac{d^3p_\chi^f d^3p_\pi^f}{E_\chi^f E_\pi^f} \right) |\mathcal{A}|^2 \delta^{(4)}(p_\chi^{in}-p_\chi^{f}-p_\pi^{f})\,;
\label{dgamma}
\eeq
with, the on-shell amplitude $\mathcal{A}\propto gv|\vec{p}_\pi^f|/m$ following from \eqref{plag}.

The energy dissipation rate can be obtained from the differential interaction rate as
\beq
\dot{E}=-\int \omega_k ~d \Gamma\,.
\label{energyrate}
\eeq 
Here, $\omega_k$ is the energy dissipated in a single phonon radiation event, $E(t)$ is the energy of the impurity as a function of time and dot denotes the time derivative, as usual.

Parameterising the initial momentum of $\chi$ as $p_\chi^{in}=\left(M+\frac{MV^2}{2},M\vec{V}\right)$, the momentum of the outgoing phonon as $p_\pi^f=\left(\omega_k, \vec{k}\right)$ and integrating over the final $\chi$-momentum, we arrive at
\beq
|\dot{E}|=\frac{g^2 n}{m^3 M^2 V} \int_{0}^{1} d{\rm cos}~\theta\int_0^{k_{\rm max}}dk ~k^3 ~\delta\left( {\rm cos }~\theta-\frac{\frac{k}{2M}+c_s}{V} \right)\,;
\label{rate1}
\eeq
where, as we have already pointed out, the description of phonons as sound waves breaks down at $k_{\rm max}=2m c_s$. Another interesting observation concerns the delta-function; in particular, because of it only $k<2M(V-c_s)$ momenta contribute to the integral. As a result, we find ourselves in an interesting position. If $k_{\rm max}> 2M(V-c_s)$ (which would imply that the perturber is lighter than the constituent of the superfluid, or that $V$ is extremely close to the sound speed), then we didn't have to be careful about the range of validity of the sound wave description. Because, we would have simply terminated the integration at $k=2M(V-c_s)$ to obtain
\beq
|\dot{E}|\propto \frac{g^2 M^2n (V-c_s)^4}{m^3 V}\,, \qquad \text{valid for} \qquad c_s<V< c_s\left(1+\frac{m}{M} \right)\,.
\label{bad}
\eeq
This is interesting only for $m\gg M$, otherwise there is not much of a velocity range for which the result is applicable. However phenomenologically limited the application, this result should capture the full energy dissipation since we didn't terminate integration by hand, it was terminated by the energy-momentum conserving delta-function well within the validity of sound wave approximation.

Now let us switch to the opposite situation, namely
\beq
V>c_s\left( 1+\frac{m}{M} \right)\,.
\eeq
This inequality is especially natural for $M>m$. In this case, if we were sloppy about the validity of the approximation at hand, we would have integrated \eqref{rate1} beyond $2m c_s$ and up to $2M(V-c_s)$. This would result into \eqref{bad} even for heavy impurity, which  would be wrong as we are about to see. In order to be consistent we should only integrate at most up to $k_{\rm max}=2m c_s$, resulting in
\beq
|\dot{E}|\propto \frac{g^2 n m c_s^4}{M^2 V}\,.
\label{goodish}
\eeq
This result captures energy dissipation in phonons with momenta up to $k_{\rm max}$, which is imposed by hand in some sense. In reality the phonons with higher momenta could be also radiated, however for them we'd have different dispersion relation, i.e. $\omega_k\neq c_s k$. Therefore, the full dissipation rate will be higher than the one given by \eqref{goodish}. Next, we will compute the dissipation rate using the more complete dispersion relation given by \eqref{disp}.

In order to proceed, we need the higher-derivative version of \eqref{plag}. Fortunately, it is easily obtainable from \eqref{nrsuplag}. Following the main theme of this letter, we are interested in the quadratic phonon action and the vertex responsible for the process $\chi\rightarrow \chi+\pi$. 
The relevant phonon Lagrangian, see the appendix for the derivation, is given by
\beq
\mathcal{L}=\frac{1}{2}\dot{\pi}^2+\frac{1}{2}\pi\Delta\left( c_s^2-\frac{\Delta}{4m^2} \right)\pi-\frac{1}{\sqrt{2}}gv\chi^2\frac{1}{\sqrt{-\Delta+4m^2c_s^2}}\dot{\pi}+\ldots\,.
\label{hdlag}
\eeq
We can see from \eqref{hdlag}, that the inclusion of the higher-spatial-derivative terms gives phonons the dispersion relation \eqref{disp}. As for the amplitude of radiating phonons by the moving impurity $\chi$, we get
\beq
\mathcal{A}\propto g v \frac{\omega_k}{\sqrt{k^2+4m^2c_s^2}}=gv \frac{k}{2m}\,.
\eeq
Here, in the last step we took into account that the radiated phonon is on-shell and made use of the explicit form of the dispersion relation \eqref{disp}. It must be emphasised that this on-shell amplitude is identical to the one of \eqref{plag}, implying the irrelevance of the high-derivative corrections to this particular on-shell vertex. However, as it can be seen from the appendix, this merely a coincidence for some of the vertices and even the vertex responsible for radiating double-phonon has nontrivial high derivative corrections.

Following the adopted procedure, we compute the energy dissipation rate as
\beq
|\dot{E}|=\int \omega_k d\Gamma\propto \frac{g^2n}{m^3M^2V}\int_{0}^{1} d{\rm cos}~\theta\int_0^{k_{\rm max}}dk ~k^3~\delta\left( {\rm cos }~\theta-\frac{\frac{k^2}{2M}+\omega_k}{kV} \right)\,,
\label{de}
\eeq
with $\omega_k$ denoting the energy of the on-shell phonon with momentum $k$. Here $k_{\rm max}$ denotes the momentum cut-off up to which the dispersion relation \eqref{disp} is valid. And, since it has been obtained in the non-relativistic limit $k\ll m$, ultimately it must be cut-off at $m$. However, we will see that because of the delta-function only $k<m$ momenta are contributing to the integral and consequently, there is no need to impose a cut-off by hand. In other words, non-relativistic perturber is  incapable of exciting a single relativistic phonon with $k>m$.

The integration in \eqref{de}  can be easily performed, resulting in
\beq
|\dot{E}|\propto\frac{g^2nk_*^4}{m^3M^2V}\,,
\eeq
with $k_*$ defined by
\beq
\frac{k_*^2}{2M}+\omega_{k_*}=k_*V\,.
\label{keq}
\eeq
In other word, $k_*$ is a maximum momentum for which ${\rm cos }~\theta\leq 1$ and therefore the delta function gives non-vanishing result in \eqref{de}. It is easy to verify that $k_*<m$, which is the reason why we have terminated the integral appropriately. The solution to \eqref{keq} is given by
\beq
k_*=\frac{2Mm}{M^2-m^2}\left( -m V+\sqrt{M^2(V^2-c_s^2)+m^2c_s^2} \right)\,.
\eeq
One can straightforwardly show that $k_*$ is non-negative independently of the hierarchy between $M$ and $m$.
The final expression for the dissipation rate simplifies in large-$M$ limit. In particular, in the leading order in $M^{-1}$ we get the following expression for the force
\beq
|F|=\frac{|\dot{E}|}{V}=\frac{g^2m n}{M^2} \frac{(V^2-c_s^2)^2}{V^2}\,.
\label{finf}
\eeq
Notice that the friction vanishes smoothly for velocities approaching the sound speed, as expected.

Important point to note is that the dissipation rates computed using the sound-wave dispersion relation, i.e. \eqref{bad} and \eqref{goodish}, significantly deviate from the correct expression \eqref{finf}. Also, as a sanity check, we can easily see that \eqref{finf} reproduces the expression for friction in case of non-self-interacting BEC in the limit $c_s\rightarrow 0$. Therefore, in the limit $V\gg c_s$ the interaction between constituents becomes irrelevant.

Let us conclude by commenting that, although we have done the calculation for a simplest possible superfluid imaginable, one should expect our statement regarding the limitations of the standard EFT for supersonic processes to apply generically, since in general the spectrum can be even richer for momenta $k>mc_s$.

\vskip 30pt

\begin{center}
{\large \bf Acknowledgements}
\end{center}
\vskip 10pt

I would like to thank Justin Khoury and Benjamin Elder for useful discussions.

\section*{Appendix: High-derivative corrections}
\renewcommand{\theequation}{A-\Roman{equation}}
\setcounter{equation}{0} 

\label{highder}
In this appendix, we derive the effective theory of phonons which will be valid for all non-relativistic momenta; i.e. we will retain the higher-spatial-derivative terms, usually dropped in the literature. The starting point is the non-relativistic Lagrangian for superfluid degrees of freedom \eqref{nrsuplag}, which we quote here for convenience
\beq
\mathcal{L}=-(\partial_j \rho)^2+2 m (v+\rho)^2\left\{ m c_s^2+\dot{\pi}-\frac{(\partial_j \pi)^2}{2 m}-\frac{g}{4m}\chi^2 \right\}-\frac{\lambda}{2}(v+\rho)^4\,,
\label{Alag1}
\eeq
In the standard derivation, the gradient term for $\rho$ is dropped followed by integrating it out from the resulting Lagrangian which becomes its algebraic function. Here, instead we retain the said gradient term and proceed with integration. The presence of the gradient term makes the procedure more involved, as we need to carry out integration order by order. Unlike the leading order phonon action, we are led to the cumbersome effective theory in the current case. Because of this, we will go after the phonon effective theory up to quartic terms. For this we need to solve for $\rho$, using its equation of motion, in terms of other fields up to quadratic order (higher order contributions to $\rho$ are irrelevant). Performing the integration, the expression of $\rho$ in terms of other fields reduces to
\beq
\rho=\frac{1}{-\Delta+4m^2 c_s^2}\left[ 2 v m \dot{\pi}-v(\partial_j\pi)^2-\frac{g}{2}v\chi^2+2m\dot{\pi}\frac{2v\mu \dot{\pi}}{-\Delta+4m^2 c_s^2}-6\frac{m^2 c_s^2}{v} \left( \frac{2vm \dot{\pi}}{-\Delta+4m^2 c_s^2} \right)^2\right]\;.~~\nonumber
\eeq
After substituting this back in \eqref{Alag1}, we get an effective theory describing the propagation and interaction of a phonon with itself and with field $\chi$. The Lagrangian consists of the following terms:

We have the quadratic phonon action
\beq
\mathcal{L}^{\pi\pi}=\dot{\pi}\left( \frac{(2vm)^2}{-\Delta+4m^2 c_s^2} \right)\dot{\pi}+\pi\Delta\pi\,.\nonumber
\eeq

The cubic interaction terms, both self- and with $\chi$, reducing to
\beq
\mathcal{L}^{\pi\pi\pi+\chi\chi\pi}=&&-2 v(\partial_j \pi)^2\frac{2vm \dot{\pi}}{-\Delta+4m^2 c_s^2}+\left( \frac{2vm \dot{\pi}}{-\Delta+4m^2 c_s^2} \right)^2 \frac{-\Delta}{-\Delta+4m^2 c_s^2} 2m\dot{\pi}\nonumber\\
&&-g v \chi^2\frac{2vm \dot{\pi}}{-\Delta+4m^2 c_s^2}\,.\nonumber
\eeq
These vertices are, for the most part, sufficient to argue the main points of this work. However, for the reference, we also quote the quartic interaction terms
\beq
\mathcal{L}^{\pi\pi\pi\pi}=v^2(\partial_j\pi)^2\frac{1}{-\Delta+4m^2 c_s^2}(\partial_j\pi)^2+(\partial_j\pi)^2\left(\frac{12m^2c_s^2}{-\Delta+4m^2 c_s^2}-1\right) \left(\frac{2 vm\dot{\pi}}{-\Delta+4m^2 c_s^2} \right)^2\nonumber\\
-4v\mu(\partial_j \pi)^2\frac{1}{-\Delta+4m^2 c_s^2}\left[\dot{\pi}\frac{2vm\dot{\pi}}{-\Delta+4m^2 c_s^2} \right]-\frac{c_s^2 m^2}{v^2}\left( \frac{2vm\dot{\pi}}{-\Delta+4m^2 c_s^2}\right)^4\nonumber \\
+\frac{1}{v^2} \left[\left( \frac{2vm\dot{\pi}}{-\Delta+4m^2 c_s^2}\right) \left( \frac{-\Delta-2m^2 c_s^2}{-\Delta+4m^2 c_s^2} \right)2vm\dot{\pi} \right]\frac{1}{-\Delta+4m^2 c_s^2}\left[\left( \frac{2vm\dot{\pi}}{-\Delta+4m^2 c_s^2}\right) \left( \frac{-\Delta-2m^2 c_s^2}{-\Delta+4m^2 c_s^2} \right)2vm\dot{\pi} \right]\,.\nonumber
\eeq
\beq
\mathcal{L}^{\chi\chi\pi\pi}=-\frac{1}{2}g\chi^2 \left( \frac{2 vm\dot{\pi}}{-\Delta+4m^2 c_s^2} \right)^2+gv^2 \chi^2 \frac{1}{-\Delta+4m^2 c_s^2}(\partial_j \pi)^2\nonumber \\
+g \chi^2 \frac{1}{-\Delta-4m^2 c_s^2}\left[\left( \frac{-2m^2 c_s^2-\Delta}{-\Delta+4m^2 c_s^2}2 v m\dot{\pi}\right)\frac{2 vm\dot{\pi}}{-\Delta+4m^2 c_s^2} \right]\,.\nonumber
\eeq
\beq
\mathcal{L}^{\chi\chi\chi\chi}=\frac{g^2}{4}v^2 \chi^2 \frac{1}{-\Delta+4m^2 c_s^2}\chi^2\,.\nonumber
\eeq

To use these relations for calculating the phonon scattering rates, it is better to canonically normalise the phonon by rescaling it as
\beq
\pi\rightarrow \frac{\sqrt{-\Delta+4m^2 c_s^2}}{2\sqrt{2}mv}\pi\,,\nonumber
\eeq
so that standard Feynman rules apply.

We would like to stressing one important point. In the limit $\lambda\rightarrow 0$, corresponding to $c_s\rightarrow 0$, there seem to be some leftover phonon self-interactions. This may give us a pause, since in this limit phonons simply correspond to free mobile massive particles. However, it can be shown that all these interactions are spurious, as they give vanishing on-shell scattering amplitudes.

We could use expressions of this appendix, to ask what happens when the $\chi$-perturber is replaced by $\Phi$. In other words, if this were the Helium superfluid then we could study the motion of a Helium atom. Once the perturber enters the superfluid, we need to revisit its nature. In the mean-field approximation, the energy-momentum of the perturber will be propagating in the form of a phonon. So the dissipation process will involve the splitting of the phonon into two phonons.

The amplitude for the relevant process ($\pi(\vec{k}_1)\rightarrow \pi(\vec{k}_2)+\pi(\vec{k}_3)$) can be readily obtained from the interaction terms we have found, and is given by\footnote{We have already put external legs of this amplitude on-shell, for simplification.}
\beq
\mathcal{A}_{\pi}=\frac{2}{v}\left\{ -2\left( \frac{\omega_{k_1}}{k_1} \frac{\omega_{k_2}}{k_2}\frac{k_3}{2m}\left(\vec{k}_1\cdot \vec{k}_2 \right)+\frac{\omega_{k_1}}{k_1} \frac{\omega_{k_3}}{k_3}\frac{k_2}{2m}\left(\vec{k}_1\cdot \vec{k}_3 \right)+\frac{\omega_{k_2}}{k_2} \frac{\omega_{k_3}}{k_3}\frac{k_1}{2m}\left(\vec{k}_2\cdot \vec{k}_3 \right)\right) \right.\nonumber \\
\left.+\frac{k_1}{2m}\frac{k_2}{2m}\frac{k_3}{2m}\left( k_1^2+k_2^2+k_3^2 \right)\right\}\,.
\eeq
The rate of this process can be computed as
\beq
d\Gamma_\pi\propto \frac{d^3k_2d^3k_3}{\omega_{k_1}\omega_{k_2}\omega_{k_3}}|\mathcal{A}_{\pi}|^2\delta^{(4)}\left(k_1^{\nu}-k_2^\nu-k_3^\nu\right)\,.
\label{phrate}
\eeq

For the purposes of demonstration we give the limiting expressions in high and low momentum limits. It is easy to show that for high momenta we get
\beq
\frac{|\mathcal{A}_\pi|^2}{\omega_{k_1}\omega_{k_2}\omega_{k_3}}\Big|_{k\gg 2m c_s}\propto \frac{\lambda^2n}{m^4}\,,
\eeq
while for low momenta, we arrive at
\beq
\frac{|\mathcal{A}_\pi|^2}{\omega_{k_1}\omega_{k_2}\omega_{k_3}}\Big|_{k\ll 2m c_s}\propto \frac{\lambda^2n}{m^4} \frac{k_1k_2(k_1-k_2)}{(m c_s)^3}\,.
\eeq
Once again, exhibiting the change in behaviour depending on how momentum transfer compares to $mc_s$

\end{document}